\documentclass[12pt]{article}

\usepackage{epsfig}

\usepackage{amssymb}
\usepackage{amsfonts}

\usepackage{color}
 
\def\be{\begin{equation}}
\def\ee{\end{equation}}
\def\ba{\begin{array}{c}}
\def\ea{\end{array}}

\def\ben{$$}
\def\een{$$}

\newcommand{\bea}{\begin{eqnarray}}
\newcommand{\eea}{\end{eqnarray}}

\newcommand{\pbr}{\prec\!\!}
\newcommand{\pkt}{\!\!\succ\,\,}
\newcommand{\kt}{\rangle}
\newcommand{\br}{\langle}

\begin{document}

\begin{center}

{\Large

Admissible perturbations and false instabilities in ${\cal
PT}-$symmetric quantum systems

}

\vspace{0.8cm}

  {\bf Miloslav Znojil}

\vspace{0.2cm}

\vspace{1mm} Nuclear Physics Institute of the CAS, Hlavn\'{\i} 130,
250 68 \v{R}e\v{z}, Czech Republic

{e-mail: znojil@ujf.cas.cz}

\end{center}

\section*{Abstract}

One of the most characteristic {\em mathematical\,} features of the
${\cal PT}-$symmetric quantum mechanics is the explicit
Hamiltonian-dependence of its physical Hilbert space of states
${\cal H}={\cal H}(H)$. Some of the most important {\em physical\,}
consequences are discussed, with emphasis on the dynamical regime in
which the system is close to phase transition. Consistent
perturbation treatment of such a regime is proposed. An illustrative
application of the innovated perturbation theory to a non-Hermitian
but ${\cal PT}-$symmetric user-friendly family of $J-$parametric
``discrete anharmonic'' quantum Hamiltonians $H=H(\vec{\lambda})$ is
provided. The models are shown to admit the standard probabilistic
interpretation if and only if the parameters remain compatible with
the reality of spectrum, $\vec{\lambda} \in {\cal D}^{(physical)}$.
In contradiction to conventional wisdom the systems are then shown
stable with respect to admissible perturbations, inside the domain
${\cal D}^{(physical)}$, even in the immediate vicinity of the
phase-transition boundaries $\partial {\cal D}^{(physical)}$.

\subsection*{Keywords}

non-Hermitian Hamiltonians; ${\cal PT}-$symmetry; boundaries of
stability; quantum catastrophes; singular perturbation theory;
elementary solvable models;

\newpage

 \noindent

\section{Introduction\label{introduction}}

The conventional textbook formulations of quantum theory
\cite{Styer} were recently complemented by several innovative
pictures of quantum world which may be characterized as a
quasi-Hermitian \cite{Geyer} {\it alias\,} ${\cal PT}-$symmetric
\cite{Carl} {\it alias\,} pseudo-Hermitian representation of quantum
mechanics \cite{ali}. Up to a few technical differences (cf. also
several more mathematical updates of the reviews in \cite{book}) all
of these methodical innovations were originally aimed at an
amendment of the description of the stable bound states in a closed
(i.e., unitary) quantum system, be it a ferromagnet \cite{Dyson} or
an atomic nucleus \cite{Jolos} or a toy-model field
\cite{BG,Milton}. All of these innovative, sophisticated treatments
of quantum reality remained compatible with the quantum theory of
textbooks. Still, the mere rearrangement of mathematical ingredients
helped to resolve several old theoretical problems, say, in
relativistic quantum mechanics \cite{aliKG} or in our understanding
of the mechanisms of the quantum phase transitions \cite{BB,Denis}.

The not quite expected user-friendliness of the new formalism (let
us call it here, for the sake of definiteness, ${\cal PT}-$symmetric
quantum mechanics, PTQM) inspired a huge parallel progress also in
multiple neighboring branches of physics. It ranged from the more
traditional quantum theory of resonances in atoms, molecules and
nuclei (and, in general, in any open quantum system, cf.
\cite{Nimrod,Ingrid}) up to the very successful modern developments
in experimental phenomenology within classical physics and, first of
all, in optics \cite{Musslimani}. Anyhow, in the present paper we
will return to the narrower, old-fashioned versions of the PTQM
philosophy, the basic ideas of which may be traced back, in
retrospective, to the Dyson's \cite{Dyson} replacement of a given,
``realistic'' Hamiltonian $\mathfrak{h}=\mathfrak{h}^\dagger$ by its
manifestly non-Hermitian isospectral partner
 \be
 H = \Omega^{-1}\mathfrak{h}\Omega \neq H^\dagger\,,
 \ \ \ \  \Omega^\dagger\Omega \neq I\,.
 \label{preco}
 \ee
The Dyson's key idea was that whenever we manage to choose the
invertible non-unitary mapping $\Omega$ as carrying a non-trivial
information about the system, we may make the ``preconditioned''
quasi-Hermitian Hamiltonian (\ref{preco}) computation-friendlier.
The price to pay seemed reasonable. In place of staying in the
conventional picture (where the prediction of the measurements is
given by the overlaps $\pbr \psi|\mathfrak{q}|\psi \pkt$ where
$|\psi\pkt$ is the wave function while symbol $\mathfrak{q}$ denotes
the observable of interest \cite{Messiah}) one merely changes the
Hilbert space (cf., e.g., \cite{SIGMA} for the more detailed
explanations) and one evaluates the analogous overlaps
 \be
 \br \psi|\Theta\,Q|\psi\kt\,,
 \ \ \ \ \ Q = \Omega^{-1}\mathfrak{q}\Omega
 \,.
 \ee
The symbol $\Theta=\Omega^\dagger\Omega=\Theta^\dagger > 0$ denotes
here the physical Hilbert-space metric \cite{Geyer} while the
time-evolution of the wave functions is assumed generated by the
non-Hermitian Hamiltonian (\ref{preco}) with real spectrum,
 \be
 {\rm i} \frac{d}{dt} |\psi\kt = H\,|\psi\kt\,.
 \label{pticy}
 \ee
In applications, the Dyson's trick proved extremely successful,
e.g., in the computational nuclear physics of heavy nuclei
\cite{Jolos}. Still, it did not inspire any immediate developments
in the abstract quantum theory itself. One of the reasons can be
seen in the manifest non-Hermiticity of Hamiltonian,
 \be
 H^\dagger = \Theta \,H\,\Theta^{-1}\,.
 \ee
Indeed, the naive use of such a ``quasi-Hermitian'' generalization
of the conventional Hermiticity opened the Pandora's box of
mathematical difficulties \cite{Dieudonne,Kato}. Most of these
questions are open or remain only partially answered at present
\cite{ATbook}.

In the context of physics, fortunately, people managed to circumvent
the danger along two alternative lines. In one of these directions
it was Bender with Boettcher \cite{BB} who noticed that for a family
of benchmark toy-model ordinary differential operator examples
 \be
 H^{(BB)}=-\frac{d^2}{dx^2} + x^2 ({\rm i}x)^\delta\,,
 \ \ \ \ \
 \delta>0
 \label{mode}
 \ee
the spectrum is real. On this ground it has been conjectured that
the Dyson's mapping $\mathfrak{h} \to H^{(BB)}$ of Eq.~(\ref{preco})
could have been inverted, returning us, in principle at least, to
the safe waters of the traditional unitary quantum mechanics. An
alternative, mathematically safer approach to the problem has been
accepted in the context of nuclear physics. The authors of review
\cite{Geyer} noticed that the major part of the formal difficulties
disappears whenever all of the eligible operators of observables are
required bounded, i.e., $H \in {\cal B} ({\cal H})$, etc.

We shall follow here the latter strategy leaving the mathematically
more complete treatment of unbounded models to mathematicians
\cite{SKbook}. Our decision will keep the necessary mathematical
manipulations on elementary level, facilitating our study of the
question of the stability of a generic ${\cal PT}-$symmetric quantum
system with respect to its small perturbations. Our assumptions will
be shown satisfied, in section~\ref{ahos}, by a family of
illustrative ${\cal PT}-$symmetric finite-dimensional Hamiltonians
$H = H(\vec{\lambda}) \in {\cal B} ({\cal H})$ taken from
Refs.~\cite{maximal,tridiagonal}. The variability of the physical
parameters of the models is confined to domain ${\cal
D}^{(physical)} \subset \mathbb{R}^J$ which is, at any integer
parameter $J=\dim {\cal D}$, {\em bounded}. The bound-state energies
are real and observable inside and only inside this domain.

The discrete anharmonic form endows our benchmark oscillators with a
remarkable phenomenological as well as formal appeal. On the formal
side, every Hamiltonian of the class will be real, asymmetric,
tridiagonal and particularly suitable for an exhaustive discussion
of stability with respect to perturbations. On the side of
phenomenology, our higher-dimensional matrices $H^{(N)}$ will be
shown interesting as an interpolation between the two extreme
scenarios, viz., between the weak- and strong-anharmonicity
dynamical regimes. In the former case one can split
$H=H^{(HO)}+H^{(perturbation)}$ and use the routine methods,
appreciating an equidistant spectrum in the diagonal unperturbed
harmonic-oscillator-like matrix $H^{(HO)}$. The much more
complicated analysis of the latter case will be of our present
interest.

For our stability-investigation purposes the choice of the
illustrative toy models initially seemed to be far from optimal. The
main reason was explained in \cite{operatorth}. Fortunately, the
removal of the obstacle (which lies in the necessity of construction
of the so called transition matrices) appeared feasible. The point
will be presented and explained in section \ref{chap3}. Another
formal merit of the model lies in the boundedness of its parametric
domains ${\cal D}^{(physical)}$. This feature (proven in
\cite{maximal}) is welcome as it makes the phase-transition
boundaries $\partial {\cal D}^{(physical)}$ of the stability of the
system experimentally as well as theoretically accessible. It is
worth adding that in mathematics, the elements of the boundary
$\partial {\cal D}$ were given the name of ``exceptional points''
(EP, \cite{Kato}). The original motivation of the introduction of
the concept of EPs appeared in perturbation theory where the domains
${\cal D}$ of parameters were usually considered complex and, in
general, unbounded (cf. {\it loc. cit.}). In the prevalent
Taylor-series form of perturbation expansions the knowledge of the
(usually, isolated) EP singularities in the complex domains of
couplings then offered a key to the rigorous determination of the
radius of the convergence of the series.

In the applied perturbation-expansion considerations the relevant
EPs were almost never real. This reflected the virtually exclusive
interest in self-adjoint Hamiltonians (see, e.g., the dedicated
special issue \cite{IJQC} from 1982 for illustration). In the
broader context of physics it soon became clear that the use of EPs,
real or complex, may go far beyond their auxiliary role in
perturbation theory. One of the best summaries of the situation was
provided during the 2010 international conference ``The Physics of
Exceptional Points'' in Stellenbosch \cite{Stellenbosch}. All of the
titles of the invited talks shared the generic clause ``Exceptional
points and \ldots''. The list of the samples and reviews of {\em
different\,} physical applications of the EP concept proved
impressive. It involved atoms and molecules in external fields,
light-matter interactions and the processes of photo-dissociation,
quantum phase transitions and the many-particle models, the
questions of stability with illustrations in magnetohydrodynamics,
the results of the experiments with the microwave billiards and
microdisk cavities plus, last but not least, the study of the
Bose-Einstein condensates and of the generic quantum phenomena
related to the spontaneous breakdown of ${\cal PT}-$symmetry.

Such a diversity enhances the importance of the EPs in physics
\cite{Heiss}. At the same time, the necessary {\it ad hoc\,}
adaptations of the related mathematics could lead to
misunderstandings. We shall restrict attention, therefore, to the
specific subset of applications in which a closed quantum system
exhibits a spontaneously unbroken ${\cal PT}-$symmetry. We shall
assume that the domain ${\cal D}$ of variable parameters as well as
its EP boundaries are real and bounded. This will enable us to
construct a strong-coupling perturbation recipe in which
 \be
 H(\vec{\lambda})=H(\vec{\lambda}^{(EP)}) + V_{(perturbation)}\,.
 \label{splitt}
 \ee
This means that the unperturbed Hamiltonian will be already
manifestly unphysical and non-diagonalizable, with its parameters
$\vec{\lambda}^{(EP)} \in
\partial {\cal D}^{(physical)}$ not lying inside the
domain of applicability of quantum theory. In section \ref{3.3} we
shall clarify the apparent contradiction by studying, in detail, the
most characteristic case of the unperturbed Hamiltonian containing
just the single Jordan block of a finite dimension. Admitting just
the finite values $K<\infty$ of this dimension we will simplify the
technicalities and we will explain the main specific features of the
resulting singular perturbation theory. We will show that even when
the unperturbed Hamiltonian has the non-diagonal, unphysical,
Jordan-block form, the evaluation of perturbation corrections
themselves remains feasible, comparatively user-friendly and fully
analogous to the more conventional degenerate versions of the
textbook Rayleigh-Schr\"{o}dinger perturbation theory.

The presentation of these methodical results will be complemented,
in section \ref{schap5}, by the explicit description of perturbation
approximations for a few low-dimensional models. After an exhaustive
analysis of these toy models living in finite-dimensional Hilbert
spaces we will be able to conclude that the quantum
unitary-evolution physics which is ``hidden'' behind the
non-Hermitian operators of observables is sound and acceptable. In
subsequent section \ref{disc} devoted to discussion we shall point
out, in particular, that the problem of the sensitivity of the
results to perturbations is highly nontrivial and that the key role
is played by a selfconsistently determined, interaction-dependent
measure $\epsilon$ of the smallness of the perturbation.

The last section \ref{chap6} is the summary. We will re-emphasize
there that one of the most important consequences of the
constructive use of the perturbation-approximation strategy near the
EP singularities should be seen in the resolution of  many puzzles
caused by the non-Hermiticity of the operators and matrices
\cite{Trefethen}. In the PTQM setting the situation is still rather
specific because the new factor which enters the game is the metric
$\Theta$. Whenever it exists, i.e., whenever the evolution remains
unitary and whenever ${\cal PT}-$symmetry remains unbroken, the
nontriviality of the metric induces an anisotropy in the physical
Hilbert space ${\cal H}$. In other words, in the PTQM setting the
physical quasi-Hermiticity constraints upon perturbations are
counterintuitive. Still, from the point of view of the control of
stability, these ``hidden Hermiticity'' constraints remain fully
analogous to the more traditional Hermiticity constraints
encountered in conventional quantum mechanics.

\section{Benchmark model\label{ahos}}

In Ref. \cite{Trefethen} one finds a number of persuasive examples
in which the knowledge of the spectrum $\sigma(H)$ offers just an
extremely poor and unreliable information about the results of the
evolution which is controlled by a manifestly non-Hermitian (i.e.,
in general, non-normal) generator $H$ in Eq.~(\ref{pticy}). In these
examples the key role is played by the so called pseudospectra
$\sigma_\epsilon (H)$. They are recommended as the main mathematical
tool of an amendment of the information. In a broad variety of
applications this tool proved able to characterize the consequences
of the generic small perturbations, i.e., the consequences of the
replacement of $H$ by the set of its perturbed versions $H+V$ such
that $V$ is small, $||V|| < \epsilon$.

The main weakness of this approach is that it is not applicable to
the PTQM models in which the Hilbert-space metric is
Hamiltonian-dependent, i.e., not only nontrivial (i.e., such that
$\Theta =\Theta (H) \neq I$) but also anisotropic (i.e., such that
its spectrum $\sigma(\Theta)$ is non-degenerate). Moreover, the
construction of the metric appears prohibitively complicated in the
prevailing majority of the examples with nontrivial pseudospectra.
This implies that the only feasible studies of the influence of
perturbations seem to be restricted to the matrix (i.e.,
finite-dimensional) models. Indeed, they trivially satisfy the
above-discussed physical (i.e., bounded-operator) constraints. In
addition, the use of the finite, parameter-dependent matrices
$H^{(N)}(\lambda)$ will also make the fundamental linear algebraic
time-independent Schr\"{o}dinger equation
 \be
 H^{(N)}(\lambda)\ |\psi_n^{(N)}(\lambda)\kt
 = E_n^{(N)}(\lambda)\ |\psi_n^{(N)}(\lambda)\kt\,,
 \ \ \ \ n = 0, 1, \ldots, N-1\,
 \label{timind}
 \ee
exactly solvable (cf., e.g., \cite{diskfx3}). In what follows, we
shall accept such a strategy of circumventing the
functional-analytic subtleties, therefore.

\subsection{Weakly non-Hermitian dynamical regime}

For illustrative purposes
we shall use the family
 \ben
 H^{(2)}(a) = \left [\begin {array}{cc} 1&a\\{}-a&-1\end
 {array}\right
 ]\,,
 \ \ \ \ \ \
 H^{(3)}(a) = \left [\begin {array}{ccc} 2&a&0\\{}-a&0
 &a\\{}0&-a&-2\end {array}\right ],
 \een
 \be
 H^{(4)}(a,b) = \left [\begin {array}{cccc}
  3&b   &0  &0\\
 -b&1   &a  &0\\
  0&-a  &-1 &b\\
  0&0&-b&-3
 \end {array}\right ]\,,
 \ \ \ \ \ \
 H^{(5)}(a,b)=\left [\begin {array}{ccccc} 4&b&0&0&0\\{}-b&
2&a&0&0\\{}0&-a&0&a&0
\\{}0&0&-a&-2&b\\{}0&0&0&-b&-
4\end {array}\right ]\,, \ldots \,
 \label{thisset}
 \ee
of the real and finite-dimensional tridiagonal matrices. This set
was introduced in Ref.~\cite{maximal}. Besides a user-friendly
nature of these non-Hermitian but ${\cal PT}-$symmetric real
toy-model Hamiltonians, their other merits lie in an enormous
phenomenological flexibility (i.e., multiparametric variability) and
in their sparse-matrix tridiagonality with an intuitive
nearest-neighbor-interaction appeal. An additional benefit is that
the set $\partial {\cal D}$ of all of the relevant phase transition
points of these models has a smooth and topologically trivial shape
of surface of a certain deformed hypercube with protruded vertices
at any matrix dimension $N<\infty$ (cf. the proof in
\cite{tridiagonal}).

From the conventional point of view all of
the weakly non-Hermitian forms of
Hamiltonians (\ref{thisset}), i.e., of the multiparametric
illustrative tridiagonal matrices
 \be
 H^{(2{J})}=\left [\begin {array}{cccc|cccc}
  2{J}-1&z&0&\ldots&&&&
  \\
  -z&\ddots&\ddots&\ddots&\vdots&&&
 \\
  0&\ddots&3&b&0&\ldots&&
 \\
 \vdots&\ddots&-b&1&a&0&\ldots&
 \\
 \hline
 &\ldots&0&-a&-1&b&0&\ldots
  \\
 &&\ldots&0&-b &-3&\ddots&
 \\
 &&&\vdots&\ddots&\ddots &\ddots&z
 \\
 &&&&\ldots&0&-z&1-2{J}\end
 {array} \right ]
 \label{sudam}
 \ee
and
 \be
 H^{(2{J}+1)}=\left [\begin {array}{ccc|c|ccc} 2{J}&z&0&0&0&0&0\\
 {}-z&
 \ddots&\ddots&0&0&0&0\\
 {}0&\ddots&2&a&0&0&0
 \\
 \hline
 {}0&0&-a&0&a&0&0
 \\
 \hline
 {}0&0&0&-a&-2&\ddots&0\\
 {}0&0&0&0&\ddots&\ddots&z
 \\
 {}0&0&0&0&0&-z&-
 2{J}\end {array}\right ]\,
 \label{levam}
 \ee
can be perceived as small and fully regular perturbations of certain
shifted and truncated toy-model harmonic oscillators having the
strictly equidistant spectrum. All of these multiparametric toy
models are eligible as the generators of the standard unitary
quantum evolution.

\subsection{Strongly non-Hermitian regime}

The energy spectra of models (\ref{sudam}) and (\ref{levam}) were
proven real and non-degenerate (i.e., in principle, observable) if
and only if the parameters lie inside a compact physical stability
domain ${\cal D}^{(J)}$ \cite{maximal}. The boundary of this domain
of stability (i.e., the hypersurface $\partial {\cal D}$) has been
shown compact. Its extreme points of maximal non-Hermiticity were
localized, non-numerically, as lying on a circumscribed hypersphere
(at odd $N=2{J}+1$) or on a prolate hyperellipsoid (at even
$N=2{J}$).

In the more detailed study \cite{tridiagonal} it has been found that
the separation of the Hamiltonians into two special cases
(\ref{sudam}) and (\ref{levam}) by the parity of their dimension $N$
is not needed. After a renumbering $ z \to g_1$, \ldots, $b \to
g_{J-1}$ and $a \to g_J$ of the couplings we introduced a redundant
parameter $t \in (0,1)$ (meaning ``time'' or ``strength of
perturbation'') and we admitted a ``slow'' (presumably, adiabatic)
variability of the couplings,
 \be
 g_n=  g_n(t)= g_n(0) \,\sqrt{\left(1-\xi_n(t) \right) }\,,
 \ \ \ \ \ \ \ \ \
 \xi_n(t) = t+t^2+\ldots+t^{J-1}+G_n t^J
 \in (0,1)\,,
 \label{lobkov}
 \ee
 \be
 g_n(0) = \sqrt{n\,(N-n)}\,, \ \ \ \ \ n = 1, 2, \ldots, J\,.
 \label{optim}
 \ee
This enabled us to specify the shape of the boundary using the
computer-assisted algebra.

During all of our considerations concerning the {\em small,
regular\,} non-Hermitian {\it alias\,} quasi-Hermitian perturbations
in Eqs.~(\ref{sudam}) and (\ref{levam}) we have to keep in mind that
our ${\cal PT}-$symmetric Hamiltonian $H$ just provides one of the
user-friendliest representations of some hypothetical, prohibitively
complicated but entirely traditional self-adjoint operator
$\mathfrak{h}$ (cf. Eq.~(\ref{preco}) or review paper~\cite{Geyer}).
Thus, our present non-Hermitian but ${\cal PT}-$symmetric
time-evolution Schr\"{o}dinger Eq.~(\ref{pticy}) must be perceived
as strictly equivalent to its hypothetical and, presumably,
untractable textbook alternatives, with the equivalence between the
two pictures determined by their mutual ``Dyson's''
mapping~(\ref{preco}). The difference between the use of $H$ and
$\mathfrak{h}$ is, in some sense, purely technical. Still, a strong
preference of the use of non-Hermitian $H$ may be recommended
whenever the non-Hermiticities become large, i.e., typically, when
the parameters get close to the EP phase-transition boundary of
quantum stability.

The strong-coupling version of models (\ref{sudam}) and
(\ref{levam}) has been found unitary (i.e., the reality of the
spectrum was guaranteed) if and only if the parameters lie inside a
physical domain ${\cal D}^{(J)}\subset \mathbb{R}^J$ \cite{maximal}.
All of the exceptional points forming the quantum phase-transition
boundary $\partial {\cal D}^{(J)}$ could have been classified by the
number $K$ of the energy levels which merge at them and,
subsequently, complexify. Thus, at $J=1$ the domain ${\cal D}^{(J)}$
is an interval (with the two energies merging at its two ends), at
$J=2$ the domain ${\cal D}^{(J)}$ is a deformed square with the
pairs of energies merging at its edges and with all four energies
merging at its four vertices. At $J=3$ we deal with a deformed cube
with protruded edges and protruded vertices, etc.

In such a visualization of the guarantee of unitarity the maximally
non-Hermitian extreme taking place at a vertex represents the merger
of the set of energy levels degenerating to a single real value
gauged, for simplicity, to zero. This is paralleled by the
convergence of every Hamiltonian in the list (\ref{thisset}) to the
respective Jordan-block-equivalent matrix
 \ben
 H^{(2)}_{(EP)} = \left [\begin {array}{cc} 1&1\\{}-1&-1
 \end {array}\right
 ]\,,
 \ \ \ \ \ \
 H^{(3)}_{(EP)} = \left [\begin {array}{ccc} 2&\sqrt{2}&0\\{}-\sqrt{2}&0
 &\sqrt{2}\\{}0&-\sqrt{2}&-2\end {array}\right ],
 \een
 \be
 H^{(4)}_{(EP)} = \left [\begin {array}{cccc}
  3&\sqrt{3}   &0  &0\\
 -\sqrt{3}&1   &2  &0\\
  0&-2  &-1 &\sqrt{3}\\
  0&0&-\sqrt{3}&-3
 \end {array}\right ]\,,
 \ \ \ \ \ \
 H^{(5)}_{(EP)}=\left [\begin {array}{ccccc} 4&2&0&0&0\\{}-2&
 2&\sqrt{6}&0&0\\{}0&-\sqrt{6}&0&\sqrt{6}&0
 \\{}0&0&-\sqrt{6}&-2&2\\{}0&0&0&-2&-
 4\end {array}\right ]\,
 \label{epthisset}
 \ee
etc. Now, we will ask what happens in the vicinity of these
strong-coupling limiting extremes under perturbations, provided that
we stay inside the physical domain ${\cal D}$ of parameters.

\section{Perturbation-independent transition matrices\label{chap3}}

Let us first recall the definition
 \be
 HQ=QS\,
 \label{proper}
 \ee
of the Jordan-block representative $S$ of a given non-Hermitian
matrix $H$. In this relation, the inter-twiner $Q$ is called
transition matrix.

\subsection{Solvable model in two dimensions\label{hus1}}

In the limit $a \to 1$ the first item in the sequence of real
matrices (\ref{thisset}) acquires, after an auxiliary shift of
spectrum $s$, the most elementary tilded form
 \be
 H^{(2)}(a) + sI
 =\widetilde{H}_0=\left[ \begin {array}{cc} -1+s&1
 \\\noalign{\medskip}-1&1+s\end {array} \right]\,.
 \label{ujb2}
 \ee
The tilde gets removed when we use prescription (\ref{proper}) and
move to the Jordan-block representative $S= Q^{-1}\widetilde{H}_0
Q=H_0$ of the unperturbed Hamiltonian, with
 \be
 Q=Q^{(2)}_{(EP)} = \left[ \begin {array}{cc} 1&1
 \\\noalign{\medskip}-1&0\end {array}
 \right]\,,
 \ \ \ \ \ \
 H_0=H^{(2)}_{0}
 =\left[ \begin {array}{cc} s&1\\\noalign{\medskip}0&s\end {array}
 \right]\,.
 \label{jb2}
 \ee
In time-independent Schr\"{o}dinger Eq.~(\ref{timind}) we now
abbreviate $|\psi_1\kt=x$ and $|\psi_2\kt=y$. For any $H=H_0+W$,
i.e., in {\em any\,} vicinity of the Jordan block $H_0^{(2)}$ we can
now assume that there exists a measure $\lambda \ll 1$ of the
smallness of the perturbation, $W_{m,n}={\cal O}(\lambda)$. Then,
the entirely general real form of the perturbation
 \be
 W= \left[ \begin {array}{cc} \alpha&\beta\\\noalign{\medskip}\gamma&
 \delta\end {array} \right]
 \label{lisy}
 \ee
can be inserted in Schr\"{o}dinger Eq.~(\ref{timind}). This makes it
equivalent to the two linear relations
 $$
  \left . \begin {array}{c}  \left( \alpha-\epsilon \right) x+ \left(
  1+
\beta \right) y=0\,,\\\noalign{\medskip}\gamma\,x+ \left(
\delta-\epsilon
 \right) y=0\,.\end {array} \right .
 $$
We eliminate $y=-{ { \left( \alpha-\epsilon \right) x}/({1+\beta}})$
from the first line, insert it in the second line and solve for the
{\em exact\,} energy,
 $$
 \epsilon_\pm =
 1/2\,\delta+1/2\,\alpha \pm 1/2\,\sqrt
 {{\delta}^{2}-2\,\delta\,\alpha+{
 \alpha}^{2}+4\,\gamma+4\,\gamma\,\beta}
 = \pm \sqrt{\gamma} + {\cal O}(\lambda)
 $$
The following conclusions are imminent.

\begin{enumerate}

\item
Models with vanishing $\gamma=0$ are less interesting, not to be
discussed immediately.

\item
For the {\em negative\,} parameters $\gamma<0$ in the perturbation
we get the {\em manifestly complex\,} energies. The unitarity and
stability will be lost. This is true for an {arbitrarily small size}
of the perturbation when measured in usual manner, i.e., in terms of
any conventional norm.

\item
For {\em positive} $\gamma>0$ the first-order corrections are real,
$\epsilon_\pm =\pm \lambda^{1/2} + {\cal O}(\lambda)$. Thus, the
parameter lies inside ${\cal D}$ and the quantum system itself
remains stable, in the leading-order approximation at least.

\item
In the higher-order computations it will be sufficient to replace
the standard Taylor-series perturbation-series ansatz by the
Puisseux series expansion in the powers of $\lambda^{1/2}$ (cf. also
Ref.~\cite{Uwe}).

\end{enumerate}

\subsection{Tridiagonal non-diagonalizable EP Hamiltonians}

The passage through a phase transition interface is equivalent to
the coincidence of parameter $\lambda$ with its exceptional-point
value $\lambda^{(EP)}$. In this limit one observes a confluence of
eigen-energies {\em and also\,} of the related eigenvectors. The
diagonalizability of the Hamiltonian is lost. A number of the
illustrative examples may be found in the Kato's book \cite{Kato}.
Let us recall that for a Hermitian Hamiltonian $H(\lambda)$ the EP
singularities $\lambda^{(EP)}$ will be complex. They will lie out of
the range of variability of the real parameter $\lambda$. From an
opposite perspective, whenever we ask for the existence of the
quantum phase transitions, i.e., for the {\em real\,} EP values
$\lambda=\lambda^{(EP)}$, the non-Hermiticity of the Hamiltonian
matrix in its vicinity becomes necessary.

The phenomenon of the EP-generated phase transition becomes
particulary interesting when the merger of the observable discrete
eigenvalues involves more than two items, i.e., say, $K \geq 2$
energy levels at once. Without the inessential spectral shift (i.e.,
with $s=0$) we may now rewrite Eq.~(\ref{proper}) as an explicit
definition of the ``unperturbed'' Jordan block $S=H_0$,
 \be
 H_0= \left [Q_{(EP)}^{(K)} \right ]^{-1}
 H_{(EP)}^{(K)}\ Q_{(EP)}^{(K)}\,.
 \ee
This formula can be perceived as an introduction of the
Jordan-block-related unperturbed basis. By construction, it will be
composed of the columns of the respective transition matrices. The
construction of these matrices was formulated as an open problem in
Ref.~\cite{operatorth}. Now, the solution can be sampled at $K=3$,
 \ben
  Q^{(3)}_{(EP)} = \left[ \begin {array}{ccc} 2&2&1
 \\\noalign{\medskip}-2\,\sqrt {2}&-
 \sqrt {2}&0\\\noalign{\medskip}2&0&0\end {array} \right]\,,
 \een
as well as at $K = 4$ and $K=5$
 \be
 Q^{(4)}_{(EP)} = \left[ \begin {array}{cccc} 6&6&3&1
 \\\noalign{\medskip}-6\,\sqrt {3
 }&-4\,\sqrt {3}&-\sqrt {3}&0\\\noalign{\medskip}6\,\sqrt
 {3}&2\,\sqrt {3}&0&0\\\noalign{\medskip}-6&0&0&0\end {array} \right]\,,
 \ \ \ \ \ \
 Q^{(5)}_{(EP)}=\left[ \begin {array}{ccccc} 24&24&12&4&1
 \\\noalign{\medskip}-48&-36&-12&-2&0
 \\\noalign{\medskip}24\,\sqrt {6}&12\,\sqrt {6}&2\,\sqrt {6}&0
 &0\\\noalign{\medskip}-48&-12&0&0&0
 \\\noalign{\medskip}24&0&0&0&0
 \end {array} \right]\,
, \ldots \,.
 \label{epth}
 \ee
It is also not too difficult to extend this construction using the
computer-assisted algebra. We shall need some of these matrices in
what follows.

\subsection{Exact solution for $K=N=3$}

In a way inspired by the second
item in (\ref{thisset})
let us consider
 \be
 \widetilde{H}_0=
 \left[ \begin {array}{ccc} -2+s&\sqrt {2}&0
 \\\noalign{\medskip}-\sqrt {2}&s&\sqrt {2}
 \\\noalign{\medskip}0&-\sqrt {2}&2+s\end {array}
 \right]
 \label{ujb3}
 \ee
as well as its Jordan block transform
 \be
 H_0=\left[ \begin {array}{ccc} s&1&0\\\noalign{\medskip}0&s&1
 \\\noalign{\medskip}0&0&s\end {array} \right]\,.
 \label{jb3}
 \ee
With an additional abbreviation of  $|\psi_3\kt=z$ we will again
take into considerations, along the same lines as above, an
arbitrary real-matrix perturbation
 \be
 W= \left[ \begin {array}{ccc} \alpha&\nu&\delta
 \\\noalign{\medskip}\mu&\beta&\sigma
 \\\noalign{\medskip}\tau&\rho&\gamma\end {array} \right]\,.
 \label{lyko}
 \ee
We shall assume that this perturbation is not too large, $W_{m,n}=
{\cal O}(\lambda)$. Schr\"{o}dinger equation then evaluates to the
three linear relations
 $$
 \left . \begin {array}{c}
  \left( \alpha-\epsilon \right) x+ \left( 1+\nu \right)
  y+\delta\,z=0\,,
 \\\noalign{\medskip}\mu\,x+ \left( \beta-
\epsilon \right) y+ \left( 1+\sigma \right)
z=0\,,\\\noalign{\medskip}\tau \,x+\rho\,y+ \left( \gamma-\epsilon
\right) z=0\,\end {array} \right .
 $$
where we eliminate
 $$
 y=-{\frac {x\alpha-x\epsilon+\delta\,z}{1+\nu}}
 $$
from the first line, insert it in the second and the third line,
eliminate
 $$
 z={\frac {x \left(
 -\mu-\mu\,\nu+\beta\,\alpha-\beta\,\epsilon-\epsilon
\,\alpha+{\epsilon}^{2} \right) }{-\delta\,\beta+\delta\,\epsilon+1+
\nu+\sigma+\sigma\,\nu}}
 $$
from the second line, insert it in the third line and, in the
normalization with $x=1$, we obtain the ultimate secular equation
  $$
  \tau\,{}-\rho\, \left( {}\alpha-{}\epsilon
  +{\frac {\delta\,{} \left( -\mu-
 \mu\,\nu+\beta\,\alpha-\beta\,\epsilon-\epsilon\,\alpha+{\epsilon}^{2}
 \right)
 }{-\delta\,\beta+\delta\,\epsilon+1+\nu+\sigma+\sigma\,\nu}}
 \right)  \left( 1+\nu \right) ^{-1}+
 $$
 $$
 +{\frac { \left( \gamma-\epsilon
 \right) {} \left(
 -\mu-\mu\,\nu+\beta\,\alpha-\beta\,\epsilon-\epsilon
 \,\alpha+{\epsilon}^{2} \right) }{-\delta\,\beta+\delta\,\epsilon+1+
 \nu+\sigma+\sigma\,\nu}} =0\,.
 $$
This equation defines the energies exactly. In the
small-perturbation regime such an equation can be reduced to its
simplified form
 $$
  \tau\,{}-\rho\, \left( {}\alpha-{}\epsilon+{{\delta\,{} \left( -\mu-
\mu\,\nu+\beta\,\alpha-\beta\,\epsilon-\epsilon\,\alpha+{\epsilon}^{2}
 \right) }}
 \right)  +
 $$
 $$+{{ \left( \gamma-\epsilon
 \right) {} \left(
 -\mu-\mu\,\nu+\beta\,\alpha-\beta\,\epsilon-\epsilon
 \,\alpha+{\epsilon}^{2} \right) }} =0\,
 $$
and, after further reduction,
 $$
  \tau\,{}-\rho\, \left( {}\alpha-{}\epsilon+{{\delta\,{}
  \left( -\mu-\beta\,\epsilon-\epsilon\,\alpha+{\epsilon}^{2}
 \right) }}
 \right)  +
 {{ \left( \gamma-\epsilon
 \right) {} \left( -\mu-\beta\,\epsilon-\epsilon
 \,\alpha+{\epsilon}^{2} \right) }} =0\,.
 $$
Assuming that $1\gg |\epsilon| \gg |\lambda|$ we get the final
formula
 $
  \tau  -\epsilon^{3}
  =0\,.$
We have a choice between three eligible small-perturbation roots of
the same size. Once we select the real one,
 \be
 \epsilon = \sqrt[3]{\tau}
 \label{tretiodm}
 \ee
our perturbative bound-state solution remains compatible with the
unitary-evolution requirement. The previous conclusions need not be
modified too much.

\begin{enumerate}

\item
The discussion of the models with vanishing $\tau=0$ will be skipped
again. They might have been discussed, if asked for, easily.

\item
For a real $\tau \neq 0$ there always exists a unique real
leading-order solution. The other two roots are complex and have to
be discarded as incompatible with the unitarity and with the
stability of the evolution.

\item
The generic corrections are proportional to $\epsilon =
\lambda^{1/3}$ so that, again, the standard Taylor-series
perturbation-series ansatz must be replaced by the Puisseux series.

\end{enumerate}

 \noindent
We have to add that both of the spurious solutions of the
perturbative Schr\"{o}dinger equation reflect the action of an
inadmissible perturbation under which the system would lose its
operational meaning. Along this line of evolution the given
Hamiltonian as well as the related physical Hilbert space would
cease to exist. There will also be no operator of metric $\Theta$.
Consequently, the metric-dependent norm of the perturbation will be
undefined. The related perturbation itself can only be characterized
as unacceptable, mathematically divergent and carrying no physical
meaning anymore.

\section{General singular perturbation theory\label{3.3}}

One of the most characteristic properties of the anomalous limits
$H(\lambda^{(EP)})$ of a generic perturbed Hamiltonian
(\ref{splitt}), with the EP parameter $\lambda^{(EP)}$ being real or
not, is that these operators cannot be diagonalized. This is a key
difference from the diagonalizable operators $H$ used in the
conventional perturbation calculations. A bridge connecting the two
areas is to be sought in relation (\ref{proper}). It characterizes
the action of $H$ in both of the diagonalizable and
non-diagonalizable cases. In the former scenario the array $Q$ of
the eigenvector columns is complete and the spectrum-representing
matrix $S$ is diagonal plus, in all of the Hermitian and
quasi-Hermitian cases, real. In the latter scenario, a key to the
search for parallels will be now sought in the low-dimensional
examples.

\subsection{Unitarity-compatible perturbations\label{ahosII}}

The general pattern of the behavior of our toy models near their EP
extremes with any $K \geq 2$ is now obvious. We can expect that in
the PTQM setting the perturbation-correction strategy will prove as
productive as in the various conventional Hermitian theories,
therefore. We also believe that its PTQM versions will be able to
clarify the essence of numerous phenomena. In the literature devoted
to the mathematics of non-Hermitian Hamiltonians one can find a
number of attempts in this direction \cite{Guard}. {\it Pars pro
toto\,} let us recall the widespread, above-mentioned conjecture
that for non-Hermitian Hamiltonians, the concept of spectrum is much
less relevant and that its use should be replaced by the more or
less purely numerical constructions of the pseudospectra
\cite{Trefethen}. We believe that in the very specific PTQM setting
such a type of scepticism is not entirely acceptable and that our
present perturbation-approximation concept could be equally useful.

Originally, the EPs only served as an insightful tool in the
conventional Hermitian perturbation theory. Recently, the massive
turn of attention of physicists to the simulations of the quantum
phase transitions by the classical-physics means in the laboratory
\cite{simulace} were also followed by the growth of interest in the
EP-related mathematics. For this reason, we believe that the
implementation of the basic ideas of perturbation theory will prove
efficient especially near the boundaries of stability of the quantum
systems exhibiting the spontaneously unbroken ${\cal PT}-$symmetry.

\subsection{Jordan block $H_0$ of any finite dimension
\label{any}}


With the vector indices running from $1$ to $K$ let us normalize
$x=|\psi_1\kt=1$, let us abbreviate $|\psi_j\kt=y_{j-1}$,
$j=2,3,\ldots,K$ and let us introduce an artificial quantity
$y_{K}=0$. The $K$ by $K$ matrix Schr\"{o}dinger equation
 $$
 (H_0+W)
 |\vec{\psi}\kt = \epsilon \, |\vec{\psi}\kt
 $$
for the $K-$dimensional ket vector $|\vec{\psi}\kt$ with Jordan
block $H_0$ (and $s=0$) may be then re-arranged into the $K$ by $K$
matrix-inversion form
 \be
 (L + Z) \vec{y}= \vec{r}\,,
 \ \ \ \ \ \
 \vec{r} = \left (
 \ba
 \epsilon - W_{1,1}\\
 -W_{2,1}\\
 \vdots\\
 -W_{K,1}
 \ea
 \right )\,,
 \ \ \ \ \ \
 \vec{y} = \left (
 \ba
 y_1\\
 y_2\\
 \vdots\\
 y_{K}
 \ea
 \right )
 \label{tarov}
 \ee
with
 \be
 L=L(\epsilon)=\left (
 \begin{array}{ccccc}
 1&0&0&\ldots&0\\
 -\epsilon&1&0&\ddots&\vdots\\
 0&-\epsilon&\ddots&\ddots&0\\
 \vdots&\ddots&\ddots&1&0\\
 0&\ldots&0&-\epsilon&1
 \ea
 \right )\,,
 \ \ \ \
 L^{-1}=L^{-1}(\epsilon)=\left (
 \begin{array}{ccccc}
 1&0&0&\ldots&0\\
 \epsilon&1&0&\ddots&\vdots\\
 \epsilon^2&\epsilon&\ddots&\ddots&0\\
 \vdots&\ddots&\ddots&1&0\\
 \epsilon^{K-1}&\ldots&\epsilon^2&\epsilon&1
 \ea
 \right )\,
 \label{9}
 \ee
and with the {\it ad hoc\,} ``shifted'' interaction matrix
 \be
 Z=\left (
 \begin{array}{cccc}
 W_{1,2}&W_{1,3}&\ldots&W_{1,K+1}\\
 W_{2,2}&W_{2,3}&\ldots&\ldots\\
 \ldots&\ldots&\ldots&\ldots\\
 W_{K,2}&W_{K,3}&\ldots&W_{K,K+1}
 \ea
 \right )\,
 \ee
representing a small ${\cal O}(\lambda)$ perturbation such that the
``artificially added'' matrix elements are trivial, $W_{j,K+1}=0$ at
all $j$.

We may easily solve Eq.~(\ref{tarov}) in closed form,
 \be
 \vec{y}= L^{-1}(\epsilon)\vec{r}-
 L^{-1}(\epsilon)Z
 L^{-1}(\epsilon)\vec{r}+ \ldots\,.
 \label{tadef}
 \ee
In a way inspired by the illustrative examples with $K\leq 4$ it
makes also sense to assume that the size of the first-order energy
correction $\epsilon$ is small but still larger than the measure
$\lambda$ of the smallness of the individual elements of the
perturbation matrix $W$ or $Z$.

In the first-order approximation it is sufficient to keep just the
first term of the right-hand side of Eq.~(\ref{tadef}). As a test,
it is instructive to check that such a solution reproduces all of
the special-case results of preceding subsections.

In both of the exact and approximate settings our vector $\vec{y}$
contains a not yet specified free parameter $\epsilon$. In fact, the
value of this parameter is not free because it is to be determined
from the last, ``artificial quantity'' constraint
 \be
 y_{K}(\epsilon)=0\,.
 \label{secula}
 \ee
Precisely such a condition was imposed, at the very beginning of our
considerations, upon the lowermost, redundant auxiliary component of
$\vec{y}$.  This is a selfconsistency condition which now plays the
role of secular equation determining all of the eligible first-order
energy corrections $\epsilon_j$ with $j = 1, 2, \ldots, K$.

\subsection{Wave-functions}

In the $K \leq 4$ illustrations of the present constructive recipe
we saw that in the scale given by $\lambda$ the magnitude of the
roots $\epsilon_j$ can vary with the disappearance of certain
elements in the matrix of perturbations $W$ or $Z$. Still, in all of
these special cases we only have to replace the conventional
Taylor-type ansatz by its Puiseux (i.e., fractional-power-series)
generalization, keeping in mind that the selection between the roots
of the secular equation is the selection between the branches of the
solution which must be required compatible with the reality (i.e.,
observability) of the energies.

In section \ref{chap3} we replaced the standard diagonalizable
unperturbed Hamiltonian by its singular Jordan-block generalization
$H_0$. Although this made the conventional Rayleigh-Schr\"{o}dinger
perturbation expansions inapplicable, their judicious replacement by
the Puiseux power series has been shown to work properly. From the
point of view of quantum physics the feasibility of such a
generalization is fortunate.

Details were displayed for the single occurrence of a
$K-$dimensional Jordan block $H_0$. The perturbation were assumed,
for the sake of simplicity, real and bounded, $W= {\cal O}
(\lambda)$. An important byproduct of our analysis was the
observation that the generic formal measure
 \be
 \epsilon_{(K)}= \lambda^{1/K}
 \label{fofo}
 \ee
of the influence of $W$ has to be often replaced by one of its
alternative versions. The reason is that formula (\ref{fofo}) only
holds when the lower leftmost matrix element of the perturbation
matrix does not vanish, $W_{K,1} \neq 0$. Otherwise, another, larger
parameter $\epsilon_{(K)}= {\cal O} (\lambda^{1/(K-1)})$ will arise
from the formalism in a way which will be sampled below. This means
that even if we guarantee that the matrix elements $W_{m,n} = {\cal
O} (\lambda)$ are all ``sufficiently small'' at all of the
subscripts, we come to the conclusion that the extent of influence
of perturbation $W$ is {\em never\,} sufficiently reliably
characterized  by the single parameter. Such an observation
extrapolates the above low-dimensional experience to any $K$. Its
general validity finds its rigorous proof in the language of section
\ref{chap3}: The leading-order truncated version of
Eq.~(\ref{tadef}) implies that, at any Jordan-block dimension $K$,
 \be
 |\psi_j\kt=\epsilon^{j-1} +{\cal O}(\epsilon^j)\,,
  \ \ \ \ \ j = 1, 2, \ldots, K\,.
  \label{hiera}
  \ee
The insertion of this estimate in secular Eq.~(\ref{secula})
reconfirms the generic validity of formula (\ref{fofo}). The result
leading to formula (\ref{hiera}) contributes to a deeper
understanding of the failure of the naive, norm-determined
perception and estimates of the influence of perturbations. This
result indicates that with the growth of the strength $\lambda$ of
the perturbation the ``unfolding'' of the wave-function components
proceeds step-by-step, in a hierarchical ordering, but in a way
which depends on the detailed matrix structure of the perturbation.

\subsection{Four-by-four matrix illustration}

The singular EP nature of our toy-model choices of $H_0$ having the
Jordan-block form will be felt by the system even if we regularize
the unperturbed Hamiltonian by its very small change and shift
inside ${\cal D}$. For all of the sufficiently strongly
non-Hermitian ${\cal PT}-$symmetric Hamiltonians $H(\lambda)$, we
can expect a survival of the necessity of an appropriate
reinterpretation of the notion of the ``sufficient smallness'' of
perturbations, reflecting the generic component-suppression pattern
(\ref{hiera}). For this reason the insight provided by the
low-dimensional special cases is still useful.

In the direct continuation of the $K\leq 3$ studies let us now turn
attention to the next, unperturbed non-diagonalizable matrix with
$K=4$,
 \be
 \widetilde{H}_0=
  \left[ \begin {array}{cccc} -3+s&\sqrt {3}&0&0\\\noalign{\medskip}-
 \sqrt {3}&-1+s&2&0\\\noalign{\medskip}0&-2&1+s&\sqrt {3}
 \\\noalign{\medskip}0&0&-\sqrt {3}&3+s\end {array} \right]\,.
 \label{ujb4}
 \ee
Once it gets replaced by its two-diagonal Jordan-block alternative
 \be
 H_0=\left[ \begin {array}{cccc} s&1&0&0\\\noalign{\medskip}0&s&1&0
 \\\noalign{\medskip}0&0&s&1\\\noalign{\medskip}0&0&0&s\end {array}
 \right]\,
 \label{jb4}
 \ee
the stability and unitarity can be studied and guaranteed, via the
specification of the admissible, non-divergent perturbations, along
the same lines as above. Still, in order to deepen our insight in
the general tendencies let us now contemplate the next entirely
general real perturbation matrix with sixteen free parameters,
 $$
 W=  \left[ \begin {array}{cccc}
 \alpha_{{1}}&\mu_{{2}}&\nu_{{3}}&\delta
 \\\noalign{\medskip}\beta_{{1}}&\alpha_{{2}}&\mu_{{3}}&\nu_{{4}}
 \\\noalign{\medskip}\gamma_{{1}}&\beta_{{2}}&\alpha_{{3}}&\mu_{{4}}
 \\\noalign{\medskip}\tau&\gamma_{{2}}&\beta_{{3}}&\alpha_{{4}}
 \end {array} \right]\,.
 $$
Having added the fourth abbreviation  $|\psi_4\kt=w$ we can reduce
Schr\"{o}dinger equation to the linear algebraic quadruplet
 $$
 \left . \begin {array}{c}
  \left( \alpha_{{1}}-\epsilon \right) x+
 \left( 1+\mu_{{2}} \right)
 y+\nu_{{3}}z+\delta\,{w}=0\,,\\\noalign{\medskip} \beta_{{1}}x+ \left(
 \alpha_{{2}}-\epsilon \right) y+ \left( 1+\mu_{{3 }} \right)
 z+\nu_{{4}}{w}=0\,,
 \\\noalign{\medskip}\gamma_{{1}}x+\beta_{{2}}y+
 \left( \alpha_{{3}}-\epsilon \right) z+ \left( 1+\mu_{{4}} \right) {w}
 =0\,,
 \\\noalign{\medskip}\tau\,x+\gamma_{{2}}y+\beta_{{3}}z
 + \left( \alpha_
 {{4}}-\epsilon \right) {w}=0\,.
 \end {array} \right .
 $$
In the normalization such that $|\psi_1\kt=x=1$ we eliminate
 $$
 y=\epsilon-\nu_{{3}}z-\delta\,{w} + {\rm higher\ order\ corrections}
 $$
and
 $$
 z={\epsilon}^{2}-\nu_{{4}}{w}+ {\rm higher\ order\ corrections}
 $$
and
 $$
 {w}={\epsilon}^{3}-\gamma_{{1}}+ {\rm higher\ order\ corrections}
 $$
and obtain the secular equation which is rather long but, in the
leading order, degenerates to the quartic polynomial in $\epsilon$,
  $$
  \tau +\gamma_1\epsilon+\gamma_2\epsilon -\epsilon^{4}=0\,.
  $$
This yields the two eligible real roots of the same size,
 \be
 \epsilon_\pm = \pm \sqrt[4]{\tau}\,.
 \label{retau}
 \ee
The spirt of the previous conclusions survives.

\begin{enumerate}

\item
The $K=4$ case is suitable for an illustration of what happens when
$\tau=0$. In such a case Eq.~(\ref{retau}) yields the single
``undecided'' option $\epsilon=0$ and  the modified leading-order
relation
 \be
 \epsilon^3=\gamma_1+\gamma_2\,.
 \label{sample3}
 \ee
The unitarity will survive when one of the three leading-order roots
remains real.

\item
For $\tau \neq 0$ there exist two real leading-order energy
corrections. The other two solutions are complex and they have to be
discarded. The corrections prove proportional to $\epsilon_\pm
\approx \lambda^{1/4}$ so that the Puisseux series ansatz should be
employed.

\end{enumerate}

 \noindent
In the less restrictive context of open quantum systems one only
needs the construction of the energies and wave functions. In the
more restrictive framework of the study of the stable quantum
systems considered in the unitary ${\cal PT}-$symmetric setting it
is necessary to select or construct also a suitable metric operator
$\Theta$. Only such an additional construction makes the theory
complete and testable. In the next section a few comments will be
made in this spirit.

\section{Acceptable, operationally defined
perturbations\label{schap5}}

The hierarchy (\ref{hiera}) of the smallness of the wave-function
corrections is an important innovative result of our singular
perturbation considerations. Still, new open questions are evoked
by our tacit assumption that the tilded-to-untilded simplification
of $H_0$ (cf. the transformation of Eq.~(\ref{ujb2}) into
Eq.~(\ref{jb2}), etc) does not really change the way of our gauging
the size of the perturbation. With an ambition of giving all of
these concepts a meaningful operational interpretation, we will now
return to the illustrative family (\ref{thisset}) of multiparametric
non-Hermitian anharmonic-oscillator-resembling $N$ by $N$
tridiagonal-matrix matrix Hamiltonians $H(\vec{\lambda})$ which
remain quasi-Hermitian in a certain well defined
$J=[N/2]-$dimensional domain ${\cal D}$ of their free real
parameters.

One of the most relevant question concerning the perturbed ${\cal
PT}-$symmetric systems now reads: How should we understand the
notion of ``small'' perturbation? We already know that the answer
will depend on whether we insist on the reality of spectrum (in the
unitary quantum theory), or not (everywhere else). The analysis is
perceivably easier in the latter case because once we follow the
Kato's book and once we interpret energies as certain analytic
functions of the parameter, we reveal that the reality of the
energies (i.e., the unitarity of the quantum evolution) is lost in
{\em any}, arbitrarily small {\em complex} vicinity of a typical
(e.g., square-root) EP singularity.

The emergence of such a paradox is not surprising because even near
the simplest (viz., square-root) EP singularity the pair of energies
$E_\pm \sim \pm \sqrt{\lambda-\lambda^{(EP)}} $ lives on a
two-sheeted Riemann surface. For the real parameters, only the
values of $\lambda> \lambda^{(EP)} $ can lie inside ${\cal D}$ and
keep the energies real. We can say that {\em no} perturbation with
$\lambda$ leaving the interior of ${\cal D}$ can be considered
small. The elementary reason is that for $\lambda \notin {\cal D}$
the necessary physical Hilbert space ${\cal H}(H)$ in which the
physics is defined {\em does not exist} at all.

For $\lambda \in {\cal D}$, the construction  of ${\cal H}(H)$ need
not be unique \cite{Geyer}. This ambiguity will be reflected by the
ambiguity of the norm of a given perturbation, leading to the
relevant differences, especially from the point of view of an
experimentalist. We can only conclude that in contrast to the
Hermitian interaction Hamiltonians, their non-Hermitian analogues
need not admit a clear and reliable separation into their
``sufficiently small'' and ``too large'' subcategories.

\subsection{$N=2$}

For our present toy models, fortunately, the specification of the
interior of ${\cal D}$ is feasible, sometimes even by purely
non-numerical means \cite{tridiagonal}. This enables us to pay the
constructive attention even to the extreme dynamical scenarios in
which the parameters coincide with, or lie close to, one of the
vertices of the EP boundary $\partial {\cal D}$. In its vicinity,
naturally, the impact of the non-Hermiticity of perturbation $V$ is
maximal \cite{maximal}. For illustration, the first, one-parametric
Hamiltonian-operator element of sequence (\ref{thisset}) can be
reparametrized in the light of Eq.~(\ref{epthisset}) and of
Eq.~(\ref{lobkov}) with optional $G_1=1$,
 \be
 H^{(2)}[a(t)] = \left [\begin {array}{cc} 1&\sqrt{1-t}
 \\{}-\sqrt{1-t}&-1\end
 {array}\right ] = H^{(2)}_{(EP)} + V^{(2)}_{(perturbation)}(t)
 \,.
 \ee
Schr\"{o}dinger Eq.~(\ref{timind}) with $\lambda = {\cal O}(t)$,
with $N=2J$, with normalization $|\psi_1^{(2)}(t)\kt =1$ and with
the single unknown wave-function component $y=|\psi_1^{(2)}(t)\kt$
leads to the two exact bound-state solutions,
 \be
 E_\pm=\pm \sqrt{t}\,,
 \ \ \ \ \ \ \ y_\pm ={\frac {-1\pm \sqrt{t}}{\sqrt {1-t}}}\,.
 \ee
This result confirms the validity of the singular perturbation
theory of paragraph \ref{hus1}. What is now new is the choice of the
special perturbation regime based on the use of single parameter
$t\in (0,1)$ measuring the strength of the perturbation. This choice
guarantees the reality of the spectrum. Inside ${\cal D}$ it
interpolates between the strong-coupling limit of the
non-diagonalizable Hamiltonian $H_0$ at $t=0$ and the weak-coupling
diagonal-Hamiltonian limit at $t=1$. This choice also determines the
perturbation matrix,
 \be
  V^{(2)}_{(perturbation)}(t)=
  \left[ \begin {array}{cc} \sqrt {1-t}-1&\sqrt {1-t}-1
 \\\noalign{\medskip}2-2\,\sqrt {1-t}&1-\sqrt {1-t}\end {array}
 \right]=\frac{t}{2}\,
 \left[ \begin {array}{rr} -1&-1
 \\\noalign{\medskip}2&1\end {array}
 \right]+{\cal O}(t^2)
 \,.
 \label{pravaporucha2}
 \ee
It is rather counterintuitive that the eigenvalues $\pm \sqrt
{2\,\sqrt {1-t}\,-2+t}$ of this unitarity-compatible interaction
term are purely imaginary. This is a peculiarity which extends also
to the higher matrix dimensions.


\subsection{$N=3$}

The second one-parametric Hamiltonian in sequence (\ref{thisset})
can be reparametrized in the same manner as above,
 \be
 H^{(3)}[a(t)] =
 \left[ \begin {array}{ccc} 2&\sqrt {2}\sqrt {1-t}&0
 \\\noalign{\medskip}-\sqrt {2}\sqrt {1-t}&0&\sqrt {2}\sqrt {1-t}
 \\\noalign{\medskip}0&-\sqrt {2}\sqrt {1-t}&-2\end {array} \right]
 = H^{(3)}_{(EP)} + V^{(3)}_{(perturbation)}(t)
 \,.
 \ee
In the domain of small $t \approx 0$ the three exact bound-state
energies
 \be
 E_0=0\,,
 \ \ \ E_\pm=\pm 2\,\sqrt{t}\,
 \label{protri}
 \ee
remain small as expected. Similar expectation is fulfilled also by
the order of smallness of the partner of Eq.~(\ref{lyko}), i.e., by
the $N=3$ analogue of Eq.~(\ref{pravaporucha2}),
 \be
  V^{(3)}_{(perturbation)}(t)
  =\frac{t}{2}\,
 \left[ \begin {array}{rrr} -2&-1&0
 \\\noalign{\medskip}4&0&-1
 \\\noalign{\medskip}0&4&2
 \end {array}
 \right]+{\cal O}(t^2)
 \,.
 \label{pravaporucha3}
 \ee
What we detect here is the disappearance of the leftmost lowest
matrix element. This means that we have $\gamma=0$ in
Eq.~(\ref{tretiodm}) so that we have to use $\epsilon \sim t^{1/2}$
rather than the generic $\epsilon_{(3)} \sim t^{1/3}$. Otherwise,
nothing really new is observed at $N=3$ since one of the energies is
a $t-$independent spectator. At the higher $J$s, for similar
reasons, one could restrict attention to the matrices of even
dimensions $N=2J$, therefore.

\subsection{$N=4$}

In the leading order approximation it is sufficient to work with the
two trivial parameters $G_n=0$ in Eq.~(\ref{lobkov}). Along the same
lines as above we obtain the exact secular equation
$9\,{t}^{2}-10\,{\epsilon}^{2}t+{\epsilon}^{4}=0$ which is solvable
in closed form,
 \be
 E_{\pm,\pm}=\pm (2 \pm 1)\,\sqrt{t}\,.
 \label{proctyri}
 \ee
Similarly, the routine evaluation of the product
 $$
 \left [Q^{(4)}_{(EP)}
 \right ]^{-1}H^{(4)}[a(t),b(t)]
 Q^{(4)}_{(EP)}
 $$
yields the $N=4$ analogue of formula (\ref{pravaporucha3}),
 \be
  V^{(4)}_{(perturbation)}(t)
  =\frac{t}{2}\,
 \left[ \begin {array}{rrrr}
 -3& -1&0&0
 \\\noalign{\medskip}6&-1&-1&0
 \\\noalign{\medskip}0&8&1&-1
 \\\noalign{\medskip}0&0&6&3
 \end {array}
 \right]+{\cal O}(t^2)
 \,.
 \label{pravaporucha4}
 \ee
Several consequences can be formulated. Firstly,  in the manner
which extends to all of the matrix dimensions $N<\infty$ the
perturbation expansions will merely contain the integer and
half-integer powers of ${t}$. This is connected with the fact that
after the transitions to the Jordan-block $H_0$, the leading-order
form of perturbation matrix $W$ remains tridiagonal.

Our choice of the illustrative anharmonic oscillators has been
fortunate in the sense that they all admit the construction of a
smooth path connecting the common,
Rayleigh-Schr\"{o}dinger-tractable weak-coupling dynamical regime
where $g_n(t) \approx 0$, i.e., $\xi_n(t)\to 1^-$, with the
extremely non-Hermitian but still safely physical strong-coupling
dynamical regime. Unambiguously, the smallness of the
unitarity-compatible perturbations was then measurable by the
smallness of our ``redundant'' parameter $t \approx \xi_n(t)
\gtrapprox 0$.

\section{Discussion\label{disc}}

One of the  most exciting features of any non-Hermitian but ${\cal
PT}-$symmetric model, quantum \cite{Carl} or non-quantum
\cite{Musslimani}, should be seen in its capability of living near,
or passing through, an instability. The description of these
processes is controlled by Schr\"{o}dinger Eq.~(\ref{pticy}) where
the non-Hermiticity of the (say, single-parameter-dependent)
Hamiltonian $H =H(t)$ enables us to modify the dynamics by a small
change of $t$. In particular, a sudden jump can occur from the
unitary evolution scenario with the real energy spectrum to a
broken-symmetry dynamical regime \cite{BB}. A complexification of
the energies then implies a sudden loss of the unitarity of the
evolution. One of the possible phase transitions {\it alias\,}
``quantum catastrophes'' \cite{catast} is encountered.

In our present paper we addressed precisely the problem of
suppression and control of the latter type of sensitivity to
perturbations. In a natural continuation of such a study one might
imagine a further continuation of this development of the theory in
two directions. In one of them the constructions would be extended
to the systems with several Jordan blocks in the Hamiltonian. This
would closely parallel our present approach and the resulting
picture would not leave the textbook framework of unitary quantum
theory of bound states. The admissible generalized, non-Hermitian
but diagonalizable Hamiltonians $H(t)$ would still be required
quasi-Hermitian in the sense of Ref.~\cite{Geyer}. After an
appropriate amendment of Hilbert space the evolution of the
quasi-Hermitian system will still be correctly reinterpreted as
unitary. The necessary innovations would be purely technical.

In the second possible branch of future developments one might admit
that the sophisticated physical Hilbert space ceased to exist.
Besides the solution of Schr\"{o}dinger Eq.~(\ref{pticy}) the new
tasks for the theoreticians will be twofold. Firstly, whenever the
energies remain real even {\em after} the phase transition, a new
Hamiltonian-dependent physical Hilbert space must be constructed
(cf., e.g., Refs.~\cite{Denis} for some comments). Secondly, in the
case of the loss of the reality of the energies, the picture of
reality cannot be based on the safe and unitary PTQM theory. Still,
the more general, non-unitary interpretations of the evolution may
enter the game as an inspiration of such a type of future research.

The latter direction of developments will be certainly less
restrictive because the spectra as well as the parameters would be
allowed complex. The non-Hermitian and non-quasi-Hermitian quantum
Hamiltonians may then describe the realistic resonant and/or open
systems. Still, such a non-quantum type of physics will use the same
mathematics. Non-Hermitian systems with non-real spectra will
exhibit various non-unitary analogues of the EP-related phase
transitions. Even the non-conservative open quantum systems will
feel the presence of the EPs. Even at a distance, without a direct
passage through the singularity, but still with the influence
reconstructed by perturbative as well as non-perturbative techniques
(cf., e.g., papers~\cite{epoints,Milburn,Doppler} for further
details and references).

In the laboratory, the manifestations of the latter types of
non-Hermiticities may be demonstrated using, e.g., the framework of
classical optics~\cite{Rueter}. Thanks to the growing number of the
experiment-oriented simulations, multiple deeply counterintuitive
EP-related phenomena may be studied. {\it Pars pro toto}, for a pair
of the wave modes with the loss and gain (simulating ${\cal
PT}-$symmetry in the medium) the authors of Ref.~\cite{Milburn}
studied the dynamics of a system which is forced to move along a
small circle circumscribing an EP singularity. They discovered and
proved that the adiabatic approximation must necessarily fail. In
their own words, ``in contrast to Hermitian systems, the dynamics
cannot be obtained by perturbative corrections to the adiabatic
prediction''. In the light of the formalism developed in our present
paper, such a conclusion might have been modified in the near
future. We believe that in similar situations the standard or
modified perturbation techniques should be admitted and used as
well. Their implementation might certainly provide a specific
insight into the structures and the dynamics of the system. At the
same time, the choices of the more sophisticated forms and features
of the Hamiltonians will certainly open multiple technical as well
as conceptual questions.

\subsection{Towards the unbounded-operator models}

In retrospective, several discoveries and rediscoveries
\cite{Caliceti-DB-Alvarez} of the unexpected robust reality of the
bound-state energy spectrum were obtained by the study of the
non-Hermitian ordinary differential Schr\"{o}dinger equations.
Unfortunately, the results were treated, for a long time, as a mere
mathematical curiosity \cite{BG,Milton-Streater}. The opinions have
only changed after Bender with Boettcher \cite{BB} noticed that such
an anomaly characterizing the manifestly non-Hermitian quantum
Hamiltonian  is valid for the whole class of the next-to-elementary
Schr\"{o}dinger equations
 \be
 -\frac{d^2}{dx^2}\psi_n(x,\delta)
 + x^2\cdot ({\rm i}x)^\delta\,\psi_n(x,\delta)
 =
 E_n(\delta)\,\psi_n(x,\delta)\,,
 \ \ \ \ \ \psi_n(x,\delta) \in L^2(\mathbb{S}_\delta)\,,
 \ \ \ \ \ n = 0, 1, \ldots\,.
 \label{BeBe}
 \ee
A few years later it was proved that whenever the exponent remains
non-negative, $\delta\geq 0$, the reality of the spectrum survives
\cite{DDT}. One must only choose $\mathbb{S}_\delta$ as a suitable
$\delta-$dependent, $\bigcap$-shaped complex contour. At the not too
large exponents $\delta<2$ one can even return to the straight real
line and choose $\mathbb{S}_\delta = \mathbb{R}$ again.

\subsubsection{Phenomenological perspective}

Naturally, the reality of the energies at the non-negative exponents
$\delta \geq 0$ seemed puzzling. It found its intuitive explanation
in the ${\cal P}-$pseudo-Hermiticity property $H^\dagger {\cal P} =
{\cal P} H$ of the Hamiltonian, with symbol ${\cal P}$ denoting the
operator of parity. For mathematicians this means that ${\cal P}$ is
the pseudo-metric in an associated Krein space \cite{Langer}. In the
context of physics the ${\cal P}-$pseudo-Hermiticity of manifestly
non-Hermitian Hamiltonians can be reinterpreted as the property of
${\cal PT}-$symmetry reflecting the mathematically equivalent
relation $H\, {\cal PT} = {\cal PT}\, H$ in which ${\cal T}$ is an
(antilinear) operator of time reversal \cite{BB}.

The ${\cal PT}-$symmetric quantum Hamiltonians in
Eq.~(\ref{BeBe}) with $\delta \geq 0$ were widely accepted as
eligible generators of unitary evolution in quantum theory
\cite{book}. For a correct probabilistic interpretation of this
process the fundamental requirement of the observability of the
energies $E_n(\delta) \in \mathbb{R}$ was only complemented by a
phenomenologically well motivated assignment of the observability
status to a charge (cf. \cite{Carl} and also the general discussion
of such a strategy in \cite{Geyer}). The introduction of the
operator of charge ${\cal C}$ led, ultimately, to a
quantum-theoretical picture in which the Hamiltonians $H=H(\delta)$
with $\delta\geq 0$ are made quasi-Hermitian
\cite{Geyer,Dieudonne} {\it alias\,} ${\cal
PCT}-$symmetric, $H^\dagger {\cal PC} = {\cal PC} H$ \cite{Carl}. In
this sense, the time-evolution associated with Eq.~(\ref{BeBe}) was
made, formally at least, unitary.

\subsubsection{Quantum physics perspective}

The unitarity is lost at $\delta<0$. The ``Hilbert-space-metric''
operator $\Theta={\cal PC}$ would cease to exist. In applications
one then has to speak about one of the best known samples of the
phase transition at $\delta=0$, better known as a spontaneous
breakdown of ${\cal PT}-$symmetry \cite{Carl}. From such a point of
view the conventional quantum harmonic oscillator with $\delta=0$
may be interpreted as unstable with respect to the perturbations
which would deform the exponent. At $\delta=0$, the quantum system
described by Eq.~(\ref{BeBe}) will be forced to perform the phase
transition of the first kind \cite{catast}. {\em Any\,} perturbation
making the exponent arbitrarily small but negative should be, in
this setting, considered, irrespectively of its norm, infinitely
large, entirely out of the scope of any consistent unitary quantum
theory.

The purely numerical nature of model (\ref{BeBe}) appeared to be one
of its increasingly serious shortcomings. Even the warmly welcome
reality of the spectrum at $\delta \geq 0$ was merely one of the
necessary conditions of the compatibility of Eq.~(\ref{BeBe}) with
its unitary-evolution interpretation. In rigorous sense, the status
of quantum model (\ref{BeBe}) is not yet fully clarified. One of the
main reasons for doubts was formulated by Siegl and
Krej\v{c}i\v{r}\'{\i}k \cite{Petr}. After a detailed mathematical
analysis the ordinary differential model~(\ref{BeBe}) was found not
to fit fully in the framework of the quasi-Hermitian quantum
mechanics (cf. also several compact reviews of the current state of
art in~\cite{book}). A new source of potential instability has been
found, at any generic $\delta>0$, in an anomalously large and next
to unpredictable influence of perturbations.

The difficulties of such a type were already predicted by
mathematicians \cite{Dieudonne}. In our present paper, the
resolution of the problem was based on Ref.~\cite{Geyer}, i.e., on
the exclusive use of bounded Hamiltonians. Without such a
constraint, one would have to clarify, systematically, all of the
relevant mathematical subtleties. A concise review of the results of
such an approach was written, recently, by Antoine and Trapani
\cite{ATbook}.

\subsection{Towards the models with complex energies
\label{chap2}}

The description of the latter irregularities has been based on the
study of pseudospectra. Their use may be expected to tame the
anomalies in the majority of the non-quantum applications of the
theory. Among them, let us mention here the discovery of the failure
of the adiabaticity hypothesis for non-Hermitian Hamiltonians
\cite{Milburn}. Recently, this discovery was complemented by a
deeper insight in \cite{Doppler}. The slightly modified team of
authors paid attention to the mode-switching in wave-guides. In a
toy-model-based analysis of the system they simulated the evolution
by a Schr\"{o}dinger-type Eq.~(\ref{pticy}) with a suitable
complex-symmetric (CS) Hamiltonian  $H^{(CS)}$ with complex
spectrum. Surprisingly enough, the authors worked with a
three-parametric family of these Hamiltonians but they merely
supported their observations by the brute-force numerical
calculations. This appeared to be one of the sources of inspiration
of our present perturbation-theory considerations. We were persuaded
that in similar analyses of the emergence of various
non-Hermiticity-related instabilities, quantum or non-quantum, the
direct use of a suitably adapted Rayleigh-Schr\"{o}dinger
perturbation-expansion technique might prove insightful and also
technically not too difficult.

We believe that in spite of our present restriction of attention to
the mere models with real spectra, one of the eligible branches of
the future study of  Hamiltonians  $H^{(CS)}$ could be based on
their EP-related split $H^{(CS)}=H_0^{(CS)}+V^{(CS)}$ where
 \be
 H_0^{(CS)}=\frac{1}{4}
 \left[ \begin {array}{cc} -2\,i{\it \gamma_1}&{\it \gamma_1}-{\it
 \gamma_2}\\\noalign{\medskip}{\it \gamma_1}-{\it \gamma_2}&-2\,i{\it
 \gamma_2}
 \end {array} \right]\,
 \label{dopp}
 \ee
would have the complex spectrum. Thus, it would still fit in our
present non-diagonalizable scenario based on the use of the
transition matrix
 $$
 Q^{(CS)}=\frac{1}{4}\left[ \begin {array}{cc} -i{\it \gamma_1}+i{\it
\gamma_2}&4\\\noalign{\medskip}{\it \gamma_1}-{\it \gamma_2}&0\end
{array}
 \right]\,
 $$
%
%
leading to the complex Jordan-block simplification of Hamiltonian
(\ref{dopp}),
 \be
 S^{(CS)}= Q^{-1}H_0^{(CS)} Q=
 \left[ \begin {array}{cc} -1/4\,i{\it \gamma_1}-1/4\,i{\it
 \gamma_2}&1
 \\\noalign{\medskip}0&-1/4\,i{\it \gamma_1}-1/4\,i{\it \gamma_2}\end
 {array}
 \right]\,.
 \label{rosa}
 \ee
Even though the spectrum is now complex, the method of perturbing
such an EP Hamiltonian remains the same.

\subsubsection{Classical physics perspective}

In classical physics and optics the perception of the unitarity is
specific, not of a central importance. Recently, the growth of
interest in the Hamiltonians with complex energies was motivated by
the growing appeal of the direct experimental relevance of the
concept of the spontaneously broken ${\cal PT}-$symmetry in
non-quantum setting. The easiness of the realization of ${\cal
PT}-$symmetry in the form of an interplay between gain and loss in
the optical and/or other media led to a boom of the study of its
multiple counterintuitive phenomenological features and consequences
\cite{Musslimani}.

In Eq.~(\ref{BeBe}), for example, infinitely many energies cease to
be real when the exponent becomes small but negative. Still, one
observes that the low-lying part of the spectrum remains real and
that there exists an infinite series of the critical negative
exponents
 $$
 \delta^{(critical)}(1)
 <\delta^{(critical)}(2)< \ldots <
 \delta^{(critical)}(M)
 < \ldots <\delta^{(critical)}(\infty)=0.
 $$
which are, precisely, the above-mentioned EP singularities. At these
points the Hamiltonian ceases to be diagonalizable. At every
subscript $M$ the emergence of the related Jordan block of dimension
$K=2$ reflects the degeneracy of the pair of energies $E_{2M-1}$ and
$E_{2M}$ which are real at $\delta>\delta^{(critical)}(M)$ and which
form a complex conjugate pair at $\delta<\delta^{(critical)}(M)$
(cf. \cite{BB} for more details).

An analogous situation is encountered with the non-diagonal Jordan
block $H_0^{(CS)}$ of Eq.~(\ref{dopp}). After its small
perturbation, one would have a tendency of forgetting about the
EP-related non-diagonalizability and of a replacement of the
non-diagonal Jordan block $H_0^{(CS)}$ by the apparently
user-friendlier conventional diagonal matrix of eigenvalues. IN this
setting the recommendation provided by our present paper is opposite
-- the unpleasant methodical discontinuity between the EP and non-EP
scenarios is to be solved in favor of the former one!

In this sense we believe that our present paper might inspire the
new perturbative studies. The constructive use of the smallness of
the perturbations has its intuitive appeal even near EPs. It might
help in the technically less straightforward non-Hermitian
scenarios, especially for the study of dynamics in the closest
vicinity of the open-system toy-model EP Hamiltonians as sampled by
Eq.~(\ref{dopp}).

\section{Summary\label{chap6}}

In the Kato's mathematical perturbation theory \cite{Kato} the
EP singularities play mainly
just the formal role of marks of the end of the
applicability of conventional weak-coupling expansions. The
approximations of the weak-coupling type will necessarily fail
near the EP
radius of convergence.
Although we
often encounter some of the most interesting physical phenomena
in such a dynamical regime, the authors
of the related papers usually call the strong-coupling
dynamics ``non-perturbative''.
Such a terminology is misleading.
There is no doubt that the
weak-coupling approximations can only be fully successful {\em
sufficiently far\,} from the natural EP boundaries.
Still,
there exist many examples of successful
perturbation recipes of a strong-coupling type.
One of them has been described and tested in our present paper.

Our interest in this problem was recently revitalized by the
adiabaticity-failure numerical studies \cite{Milburn,Doppler} as
well as by the papers in which the existence of the
non-Hermiticity-related instabilities was deduced using the {\it ad
hoc\,} concept of the pseudospectrum \cite{Trefethen,Petr,Krej}. In
this context we imagined that all of the similar identifications of
the instabilities contain an internal contradiction because these
identifications are made, exclusively, in the auxiliary Hilbert
spaces in which neither the parametric domain ${\cal D}$ nor the
physical metric operator $\Theta \neq I$ are properly taken into
consideration. Thus, in spite of the well known fact that the
simplification $\Theta  \to I$ is often admissible in non-quantum
calculations, the difficulty of the quantum-theoretical necessity of
the construction of $\Theta \neq I$ is often being circumvented
rather than identified as the main task.

In the literature the omission of the correct account of the
anisotropic $\Theta$s is often accompanied by the absence of the
attention paid to the ``methodical discontinuity'' between the {\em
non-diagonalizable nature} of the strong-coupling EP limit $H_0$ and
the {\em diagonalizability} of any Hamiltonian defined inside the
domain ${\cal D}$, i.e., after the strong-coupling perturbation is
being turned on. In our present paper we tried to remove a mental
barrier by having kept the unperturbed Hamiltonian non-diagonal.
Even after the inclusion of the perturbation we kept the transition
matrices $Q$ unchanged.

The trick of having {\em the same} unperturbed basis before and
after the perturbation has been shown to work well. We showed that
it converted our perturbed-EP versions of the Schr\"{o}dinger
equation, via an appropriate perturbation-expansion ansatz, into an
order-by-order solvable problem. The resulting non-Hermitian
strong-coupling formalism acquired a self-consistent but still
explicit-construction nature.

The main new mathematical feature of the whole proposal may be seen
in the fact that the ``measure of the smallness'' $\epsilon$ of the
separate perturbations has the form which can vary, first of all,
with the size and sign of the individual matrix elements of the
perturbation in question. In addition, the form of the perturbation
approximations has been revealed to vary also with the dimension $K$
of the particular unperturbed Jordan block $H_0$. This formed a
self-consistency pattern: The value of the expansion parameter
$\epsilon$ has been found to coincide with a root of a
perturbation-dependent secular-like polynomial of the $K-$th order.
In parallel, parameter $\epsilon$ has been shown to order the
wave-function components, making them arranged in a
``step-by-step-unfolding'' hierarchy (cf. Eq.~(\ref{hiera})).

In the language of physics our perturbation-approximation recipe
puts several known facts into an entirely new perspective. This was
illustrated via our family of ${\cal PT}-$symmetric $N$ by $N$
matrix Hamiltonians of Eqs.~(\ref{sudam}) and (\ref{levam}). The
list of their remarkable features as availaable in
Refs.~\cite{maximal,tridiagonal} (and including their capability of
a $t-$parametrized interpolation between the weak- and
strong-coupling dynamical extremes) was complement here by several
new items. The most remarkable one of them is seen in an easy
constructive tractability of perturbations in the strong-coupling
dynamical regime. Our models also exhibited a user-friendliness in
the secular polynomial context. The ``generic'' $K-$th-root value
(\ref{fofo}) of the perturbation-weighting parameter
$\epsilon_{(K)}$ degenerated, for our admissible anharmonic
oscillator and due to the specific tridiagonal-matrix structure of
perturbations $V$, to the mere square-root expression given by
Eq.~(\ref{protri}) at $K=3$ and by Eq.~(\ref{proctyri}) at $K=4$.

We may summarize that our illustrative non-Hermitian toy models
(\ref{sudam}) and (\ref{levam}) are sampling, in an almost optimal
manner, the ways of suppression of the various forms of a
phase-transition-onset instability. Counterintuitive as such
resistance against perturbations may seem, its mechanism has been
clarified as caused by the consequent use of the specific EP
unfolding. Among all of the eligible generic energy-correction roots
$\epsilon_{(K)}\sim \sqrt[K]{\lambda}$ of the degeneracy-removing
secular Eq.~(\ref{fofo}), just one or two had to be selected. In
other words, one has to keep in mind the fact that from the point of
view of physics the perturbation can only be realized in the {\em
existing} physical Hilbert space which is characterized by a {\em
non-singular} inner-product metric $\Theta$. From this perspective,
the apparently high sensitivity of the perturbed spectrum to certain
subtle details of the form of the perturbation should be understood
as a mere mathematical artifact which we have, in our
strong-coupling perturbation approach, fully under control.


\section*{Acknowledgement}

The work was supported by the GA\v{C}R Grant Nr. 16-22945S.


\end{document}